\documentclass[12pt,preprint]{aastex}

\newcommand{\kms}{{\rm\ km\ s^{-1}}}
\newcommand{\cm}{{\rm\ cm}}

\newcommand{\myemail}{sha@kias.re.kr}

\slugcomment{Submitted to ApJL}

\shorttitle{Singly-Peaked Ly$\alpha$ from Starbursts}
\shortauthors{Sang-Hyeon Ahn}
\received{4 Aug 2003}
\accepted{.. .. 2003}
\begin{document}

\title{Singly-Peaked P-Cygni type Ly$\alpha$ from starburst galaxies}

\author{Sang-Hyeon Ahn}
\affil{School of Physics, Korea Institute for Advanced Study,\\
207-43 Cheongnyangri-dong, Dongdaemun-gu, Seoul, 130-722, Korea}
\email{\myemail}

\begin{abstract}
We present results of Monte Carlo calculations for the Ly$\alpha$ line 
transfer in an expanding dusty supershell, where Ly$\alpha$
source is a well-localized star cluster in a starburst galaxy.
The escape of Ly$\alpha$ photons
from such system is achieved by a number of back-scattering, and
so a series of emission peaks are formed redward of the systemic
redshift by back-scattering. However, majority of observed 
Ly$\alpha$ emission from starbursts show singly-peaked 
asymmetric profiles. We find in this paper that, 
in order to form a singly-peaked Ly$\alpha$ emission, 
dust should be distributed in the ionized bubble, 
as well as within the supershell of neutral hydrogen.
We also find that the overall escape fraction of Ly$\alpha$ photons
is determined by the HI column density of the supershell,
the expansion velocity of the supershell, and the spatial distribution 
of dust. However, the kinematic information of the expanding supershell
is preserved in the profile of Ly$\alpha$ emission even when
the supershell is dusty. Our results are potentially useful 
to fit the P-Cygni type Ly$\alpha$ line profiles from starburst 
galaxies, either nearby galaxies or high-$z$ Lyman break galaxies (LBGs).
\end{abstract}

\section{Introduction}

Ly$\alpha$ is the most prominent line feature in the rest-frame
ultraviolet (UV) spectra of starburst galaxies, and so Ly$\alpha$ is
often used as a redshift indicator. Recently a large number of 
star-forming galaxies have been spectroscopically observed,
either by using the Lyman break method \citep{ste96,ste99,sha03}
or by using gravitational lens \citep{fra97,fry02}. The detailed 
review was given by \citet{tan03}. The rest-frame UV spectra of 
those galaxies often show unique emission.
In these cases, we usually identify this emission as Ly$\alpha$,
and use them as a redshift indicator. Moreover, we propose
that this unique feature has more sources of astrophysical information 
of those star-forming galaxies.

Ly$\alpha$ profiles of starburst galaxies can be classified into
three types: P-Cygni type emission, broad absorption,
and symmetric emission. The P-Cygni type emission consists
of absorption in red part and the asymmetric emission. However, 
the rest-frame UV continua of high-$z$ galaxies are often too weak
to be detected, and so we can only see the asymmetric emission
lines. The fraction of symmetric emission is
small when compared with the other two types, and thus
we will not go into detail of the symmetric type.

\citet{kun98} and \citet{sha03} showed that the kinematics 
of ambient material is crucial to determine whether emergent Ly$\alpha$ 
has either broad absorption or P-Cygni type emission.
P-Cygni type Ly$\alpha$ emission can be seen only in those galaxies
whose interstellar absorption lines are blueshifted with respect to 
either Ly$\alpha$ emission or stellar atmospheric lines. 
It seems to be a general consensus that outflowing motion is 
caused by either galactic superwind or galactic supershell 
\citep{la98,hec98,tan01,al02,all03a,all03b,sil03,spi03}.
The formation of asymmetric Ly$\alpha$ by outflows was discussed
qualitatively by \citet{ten99}. 

We study in this paper
the formation of P-Cygni type Ly$\alpha$.
Ly$\alpha$ radiative transfer is mainly achieved
by back-scattering processes in outflowing media. 
We have studied the role of back-scattering processes 
in the formation of Ly$\alpha$ line profile \citep{all03a,all03b}.
For the case of dust-free supershell, we found that a series of peaks
appear in the red part of Ly$\alpha$. In that paper we emphasized
that the kinematic motion is imprinted on both the width of each peaks
and the velocity difference between peaks. However, majority of the
observed Ly$\alpha$ emission lines are dominantly singly peaked.
Possibly we can attribute this discrepancy to the previously neglected
dust extinction in the radiative transfer. 
We study in this paper the Ly$\alpha$ line formation mechanism by
interplay among the Ly$\alpha$ resonance scattering
in optically thick supershell of neutral hydrogen,
the spatial extension of dust in the HII bubble, and the
kinematics of the supershell with respect to Ly$\alpha$ sources
at the center of the supershell. 

\section{Model}

Galactic supershells are very well-known structures in the nearby 
starburst galaxies \citep{mar95,mar98,kot02}.
Their origin is known to be multiple explosions of
the supernovae in active star-forming regions. In this paper we adopt
a model galaxy of high-$z$ starburst galaxies in which Ly$\alpha$ sources 
are presumed to be located at a restricted part of the galaxy
and surrounded by a dusty galactic supershell. The supershell is assumed
to be made of uniform medium of neutral hydrogen and expanding in a bulk manner.
We assume that the interior of the supershell is fully ionized, 
which is vacuum in the sense of Ly$\alpha$ scattering.
The model in this study is very simple, and it has several model
parameters. However, the main purpose of this paper is to emphasize 
the role of spatial distribution of dust in destroying the secondary
and higher peaks. We will itemize and clarify the assumptions in this 
study as follows.

\begin{enumerate}
\item 
The starbursting region is assumed to be localized in a region of 
the model galaxy. This assumption is based on the fact that almost all the
observed Ly$\alpha$ emission lines are asymmetric. If the galaxy were 
rotating and its star-forming regions are scattered along spiral arms,
we might see symmetric Ly$\alpha$ emission lines.
Hence we can consider that the star-forming regions are localized 
in a rather narrow region in the galaxies.
The HST images of about half of the LBGs show
the amorphous shape of those high-$z$ galaxies, while the other half 
show the nuclear starburst \citep{cal01}.
Here we assume that even in the amorphous galaxies, only the brightest 
star-forming region is a dominant source of Ly$\alpha$ photons.

\item
The ionized inner part of the supershell can also be a source of 
Ly$\alpha$ photons. However, the fraction of photons, generated 
in the ionized part of the supershell, is smaller 
than that of the stellar Ly$\alpha$ photons \citep{ten99}.
Moreover, there is no kinematic difference between the photons 
back-scattered by the neutral supershell and the photons emitted 
by ionized supershell. Hence we will neglect photons emitted
from the ionized inner part of the supershell. 

\item
We adopt a typical width of input Ly$\alpha$ profile 
$\sigma_{\rm Ly\alpha}=50\kms$ in this paper.
\citet{leg97} observed H$\alpha$ emission of a nearby
HII galaxy Haro 2. The H$\alpha$ line width is $\sigma_{\rm H\alpha}=50\kms$ 
which is equivalent to $\sigma_{\rm Ly\alpha}=50\kms$.
The input line width of Ly$\alpha$ affects the width of emergent
emission peaks. However, it ranges between $50$ and $115\kms$ 
\citep{pet01}, which justifies our choice.

\item
The continuum is neglected in this paper. 
According to \citet{sha03}, the Ly$\alpha$ profiles for $38\%$
of their sample of LBGs consist of a combination of both emission
and absorption. The equivalent width (EW) of broad absorption
is $10-20{\rm \AA}$, which will be dumped in the red wing of
the primary peak. However, those photons must be partially destroyed
by dust extinction. Moreover, dust-free galaxies
would have Ly$\alpha$ equivalent widths of $50-200{\rm \AA}$
\citep{cha93}, which is much larger than the absorption 
EW of the continuum. Hence we assume that the continuum will 
give a minor contribution.

\item
We also assume that the galactic supershell fully 
covers the starburst region, and that $R_{min}=0.9R_{max}$.
Here $R_{max}$ is the outer radius of the supershell and
$R_{min}$ is the inner radius of the supershell.

\item
We also assume that the HI column density outside 
the supershell is negligible, because the supershell 
is assumed to be of galactic scale. 
We adopt the column density 
of neutral hydrogen $N_{\rm HI}=10^{19}-10^{20} {\rm cm}^{-2}$ 
in accordance with Voigt fittings done by \citet{kun98}.
This column density corresponds to the line-center optical depth by
$\tau_0 \equiv 2.27\ \left({b / 80\kms}\right)^{-1}
\left[{N_{\rm HI} / {10^{14} \rm cm^{-2}}}\right]$.
Here $b$ is the Doppler parameter corresponding to turbulence,
which is typically $80\kms$ for nearby HII galaxies \citep{kun98}.
This gives us the Voigt parameter $a = 7.60\times 10^{-5} (80\kms/b)$.

\item
We also assumed for simplicity that the neutral hydrogen in
the supershell is uniformly distributed. In fact, the supershell 
can either be patchy or have density gradient, which is also important 
for escape of Lyman limit photons. 
The geometry of the supershell is assumed to be spherical for
simplicity. The supershell evolves more rapidly toward the steepest
density gradient \citep{sil98}, but it evolves to be of galactic 
scale. Thus we can assume that the supershell is spherical.
However, more work for complicated geometry can be expected
in future work.

\item We adopt the typical expansion velocity of 
the supershell $V_{exp}=200\kms$. According to \citet{kun98},
the neutral gas in nearby starburst galaxies
along the line of sight is being pushed by an expanding envelope 
around the HII region, outflowing at velocities close to $200\kms$.
Moreover, in the composite spectrum of 
high-$z$ galaxies, \citet{sha03} measured an average
blueshift for the strong low-ionization interstellar features
of $\Delta v=-150\pm60\kms$ with respect to the stellar systemic 
redshift. Thus we adopt $V_{exp}=200\kms$ as a typical value. 

\item 
The bulk expansion is assumed, although
there can be radial velocity gradient in the supershell.
The real supershell must be complex, but we restrict the current
scope of study within cases of bulk expansion. Here the bulk 
expansion implies that dust and gas are also strongly coupled.

\item
We assume that dust is distributed uniformly in the ionized bubble,
as well as within the supershell.
When we let $R_d$ be the inner radius of the dust shell,
$R_d$ can be less than $R_{min}$. The outer radius of dust shell 
is assumed to be equal to $R_{max}$.
Although the dust size distribution and the number density
are surely altered by the shock waves, we assume in this paper 
that they are same both in the ionized bubble and within the supershell.

\item 
We defined a radial dust opacity through the dust shell $\tau_d$,
in order to reconcile with the usual way of estimating dust 
opacity by analyzing the extinction of the UV continuum from central stars
in the central region of HII region.
Another important physical parameter related with dust
abundance is dust-to-gas ratio. In particular,
we define the dust-to-HI ratio in the neutral supershell
by ${\bf D}=\tau_d/ \{N_{\rm HI} \sigma_d (1-A)\}
\{(R_{max}-R_{min})/(R_{max}-R_d)\} m_d/m_{\rm HI}$.
Here the typical mass of grains is defined by $m_d=4\pi/3 a_d^3 \rho_d$,
$\sigma_d$ is the absorption cross-section of dust grains,
$A$ is albedo, $m_{\rm HI}$ is the mass of hydrogen atom, 
and $a_d$ is the radius of dust grains effective at $1216{\rm \AA}$. 
We adopt the mass density of dust grains $\rho_d= 1 {\rm g}\cm^{-3}$ 
and the size of dust grain $a_d=1.9\times10^{-6}\cm$.
We adopt the typical albedo $A=0.5$ in accordance with \citet{dra84}.

\end{enumerate}

We use the Monte Carlo method whose detailed description was presented
in our previous papers \citep{all00,all01,all02,all03a}.
In the Monte Carlo code, we calculate the integrated path lengths 
of a Ly$\alpha$ photon before its escape only if the photon is passing 
through the dust shell.
Thus the integrated dust opacity experienced by the photon is given by
$\tau_d^e = {\tau_d / \tau_0} \sum_{i} S_i$.
Here $S_i$ is a path length in units of $\tau_0$ 
along which a Ly$\alpha$ photon propagates
between the $(i-1)$-th scattering and the $i$-th scattering 
within the dust shell in $R_d<r<R_{max}$. A more detailed description of 
the Monte Carlo code for dust-free supershell was given in \citet{all03b}.

\section{Results}

We present the emergent profiles which are calculated using the Monte
Carlo method in Figures 1 and 2. We can see that there are 
two peaks for each case, which is similar to our previous study 
\citep{all03a,all03b}. We call the peaks
at $V\simeq V_{exp}$ the primary peaks, and the peaks at
$V\approx 3V_{exp}$ the secondary peaks.
Figure 1 show the emergent profiles for $N_{\rm HI}=10^{19}\cm^{-2}$
and $N_{\rm HI}=10^{20}\cm^{-2}$.
When we compare the left panels with $R_d=0.4R_{max}$
to the right panels with $R_d=0.9R_{max}$, we can see that
the secondary peaks are destroyed efficiently if the dust shell
extends into the ionized bubble (i.e. cases with $R_d=0.4R_{max}$). 

This fact means that the emergent Ly$\alpha$ photons, forming
the peaks, have experienced almost the same amount of dust extinction,
if dust is distributed only within the supershell.
This can be explained as follows. Every back-scattering causes redshift
of Ly$\alpha$ photons. Since the supershell is optically very 
thick, Ly$\alpha$ photons are back-scattered at the shallow inner
part of the supershell. Hence, the dust opacity contributed
by path length within the supershell is relatively small, and 
the contribution during the final escape dominates the total 
dust opacity. Hence, the total dust opacity for most of Ly$\alpha$ 
photons escaped after back-scattering becomes similar, which results 
in the inefficient decrease of secondary peaks.

The situation becomes different when the dust shell extends 
into the ionized bubble, $R_d < R_{min}$.
While they are scattered back and forth in the bubble by back-scattering,
the Ly$\alpha$ photons are redshifted and eventually escape to form
the secondary and the tertiary peaks. Thus, back-scattered Ly$\alpha$ 
photons have large path length within the dust shell in the ionized 
bubble. The dust opacity contributed by the dust shell 
within the ionized bubble is roughly proportional to the number of 
back-scattering. Hence the secondary and the tertiary peaks are 
more rapidly decreased than the primary peak.

The survival fraction of Ly$\alpha$ photons
is mostly affected by the dust opacity within the supershell,
which is determined by two factors; one is the dust opacity
within the supershell, and the other is the integrated path length
within the supershell. According to theoretical 
researches \citep{ada72,har73,neu90,all02},
we usually call the medium extremely thick when $a\tau_0 \ge 10^3$. 
In this case, a Ly$\alpha$ photon experiences a large number of resonance 
core scattering, during which it happens to scatter with a hydrogen 
atom fastly moving at the tails of the Boltzmann velocity distribution.
Then, the photon becomes a wing photon, for which the medium
is transparent. The wing photon restores gradually to the core photon
while it experiences a series of wing scattering, during which
the photon wanders around the medium. This wandering can be regarded
as a spatial diffusion, and can be described by random walk process.
The photon has now core frequency again, and it repeats
such processes. We call this one cycle an excursion.
During such an excursion, Ly$\alpha$ can escape the medium.
As a corollary, Ly$\alpha$ photons suffer a larger amount of dust
extinction either when the integrated path length of those photons
within the supershell is increased or when the dust opacity within 
the supershell is very high. 

Firstly, the dust opacity within the supershell is
$\tau_d (R_{max}-R_{min})/(R_{max}-R_d)$. Thus, although the
HI column densities are same, the dust opacity
within the supershell increases as $R_d$ approaches to $R_{min}$.
As a result, in Figure 3, the survival fraction 
curves for $R_d=0.9R_{max}$ are steeper than those for $R_d=0.4R_{max}$
for cases with the same $N_{\rm HI}$ and $b$.

Secondly, as $a\tau_0$ increases, both the number of wing scattering per
excursions and the number of excursions within the supershell increases.
Hence, the path length within the supershell is an increasing function 
of $a\tau_0$. In Figure 3, for cases with the same Doppler 
parameter $b=80\kms$, we can see that the curve 
with $N_{\rm HI}=10^{20} \cm^{-1}$ is decreased much faster than 
that with $N_{\rm HI}=10^{19} \cm^{-1}$.

Thirdly, we know in Section 2 that that $a \propto b^{-1}$ and 
$\tau_0 \propto b^{-1}$. Thus, $a\tau_0 \propto b^{-2}$. 
Since the number of excursion within the supershell is 
proportional to $a\tau_0$. We can expect that 
the destruction of the secondary and higher peaks is enhanced
for small $b$. We show in Figure 2 the changes of profiles for different
Doppler widths, when both the column density of neutral 
hydrogen $N_{\rm HI}$ and the dust configuration 
(or $R_d$ and $\tau_d$) are same. The secondary peak is decreased more
rapidly as $b$ is decreased. In addition, the width of the primary
peaks decreases for small $b$. The red part of the primary peak is 
formed of the photons which are scattered backward.
Those photons experiences dust extinction both within the supershell
by resonance scattering and in the dust shell. Thus they can be easily 
destroyed by dust during those scattering processes, and so 
the primary peak becomes narrow. We show in Figure 3 the variation of 
survival fraction of Ly$\alpha$ photons for the cases 
of different Doppler parameters, while $N_{\rm HI}$, 
$\tau_d$, and $R_d$ are same. When we compare the solid line with solid 
dots with the dotted line with solid dots, we can see that the curve 
for cases with small $b$ decreases rapidly.

On the other hand, we can see in Figures 1 and 2 that both the widths of 
peaks and the velocity difference between peaks are largely 
insensitive to dust opacity.
We showed that the kinematics of outflowing material
is imprinted on both the velocity width of the peaks
and the velocity differences between peaks \citep{all03a,all03b}.
Therefore, those characteristics can be used potentially to estimate 
the expansion velocity of the supershell even in dusty interstellar 
media.

\section{Summary}

% summary
In this paper we emphasize that the spatial distribution
of dust grains in the supershell is as important in forming
the P-Cygni type Ly$\alpha$ emission profiles as the kinematics of the
supershell. The escape probability of Ly$\alpha$ photons
is greatly enhanced when the supershell is expanding
with respect to the central Ly$\alpha$ sources.
That is, the Ly$\alpha$ resonance scattering is off-centered 
and the escape probability of Ly$\alpha$ photons is enhanced. 
As the number of back-scattering increases, Ly$\alpha$ photons
are redshifted more and more, which enhances the escape probability.
Ly$\alpha$ photons are scattered backward by high HI column density of 
expanding supershell. Therefore, in the cases of dust-free supershell, 
these back-scattering processes result in a series of peaks in the redward 
of Ly$\alpha$. However, majority of observed Ly$\alpha$ emission
show no such peaks. In this paper, we have found that the single peak
can be formed, only when dust extends in the ionized bubble, 
as well as within the supershell.
In this case, the integrated path length in dust shell
increases as Ly$\alpha$ photons scatter back and forth
by the shallow part of the inner wall of the supershell.
Thus, the integrated dust opacities increases as the inner radius 
of the dust shell decreases, which is main discovery of this study.
Moreover, when Ly$\alpha$ photons suffer a large number of
resonance core scattering and wing scattering
within the supershell, their path lengths become very large.
Thus, the dust opacity within the supershell is also one of decisive
factor that determines the overall escapability of Ly$\alpha$ photons.
Furthermore, both the optical depth and the Doppler
parameter of the supershell are another crucial factor
to determine the survival fraction of Ly$\alpha$ photons.
We have found that the kinematic information of
outflowing media surrounding Ly$\alpha$ sources is imprinted
on the Ly$\alpha$ profiles, and it is conserved even when the medium
is dusty. However, our model is very simplified one, whose
limitation is described in Section 2 in detail.
More work is needed in the future.

% Figure 1
\begin{figure}
\plotone{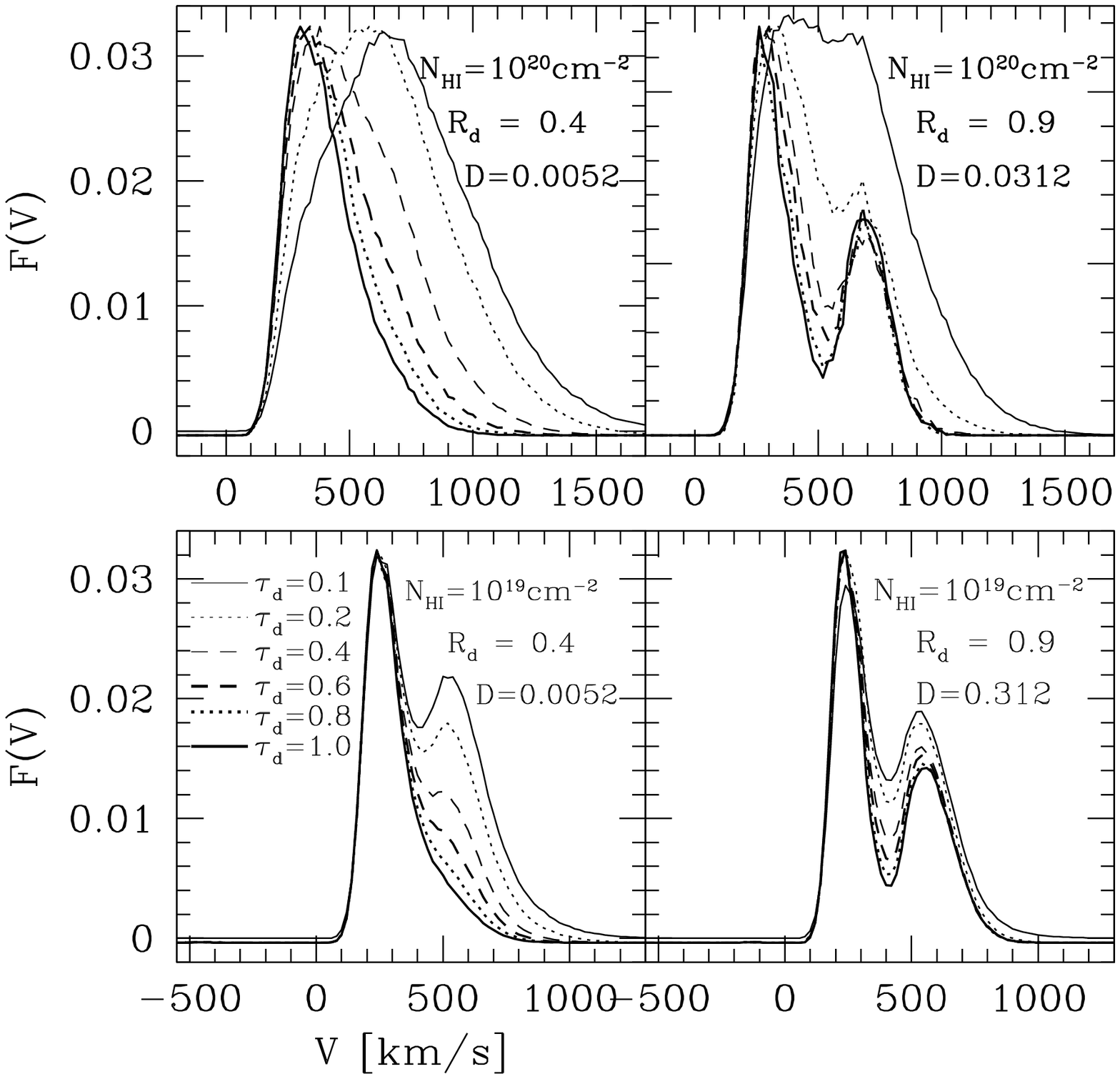}
\caption{Profiles of emergent Ly$\alpha$ emission. The upper panel is
for cases with $N_{\rm HI}=10^{20}\cm^{-2}$, and the lower panel is 
for cases with $N_{\rm HI}=10^{19}\cm^{-2}$. The inner radii of cases 
in the left panels are $0.4R_{max}$, and those of left panels 
are $0.9R_{max}$. The expansion velocity of the supershell $V_{exp}=200\kms$,
and the Doppler parameter $b=80\kms$ in common. In order to show 
the relative flux ratio between the primary and the secondary peaks, 
all the profiles are normalized with respect to the primary peak. 
The vertical coordinate means flux, and the survival fraction
of Ly$\alpha$ photons is shown in Figure 3. The dust-to-gas ratios 
in the supershell for $\tau_d=1$ are given in each box. 
Note that the dust-to-(HI+He) ratios of Our Galaxy and the Large 
Magellanic Cloud are 0.007 and 0.002, respectively (Whittet 1992).}
\label{fig1}
\end{figure}

% Figure 2
\begin{figure}
\plotone{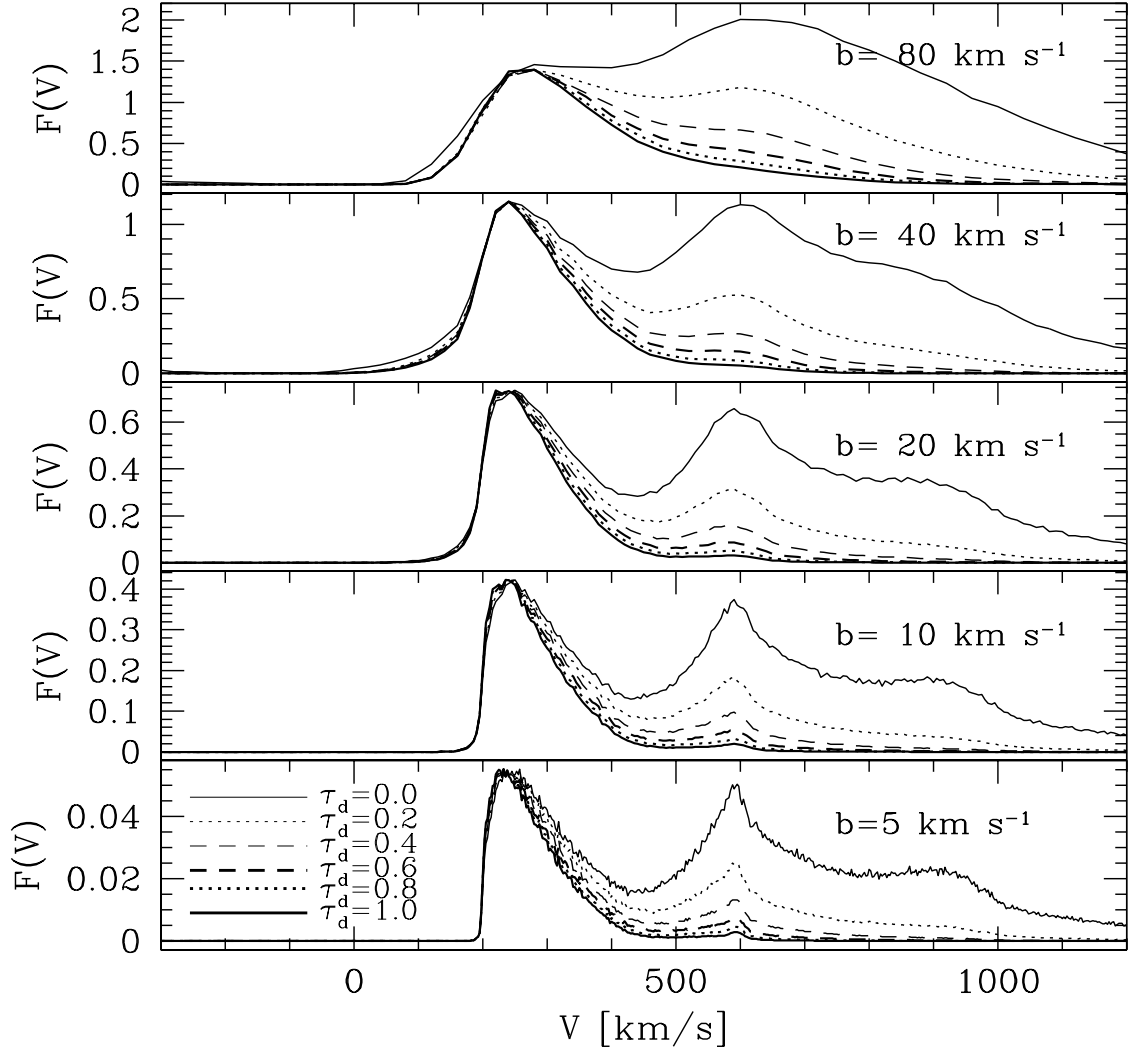}
\caption{Profiles of emergent Ly$\alpha$ emission for cases with
different Doppler widths and same other parameters 
($N_{\rm HI}=10^{19}\cm^{-2}$, $V_{exp}=200\kms$, $R_d=0.4R_{max}$). 
All the profiles are normalized with respect to the primary peak.
The relative strength is represented by dotted lines of survival 
fraction curves in Figure 3. Note that the secondary peaks decreases 
more sensitively as $b$ decreases.}
\label{fig2}
\end{figure}

% Figure 3
\begin{figure}
\plotone{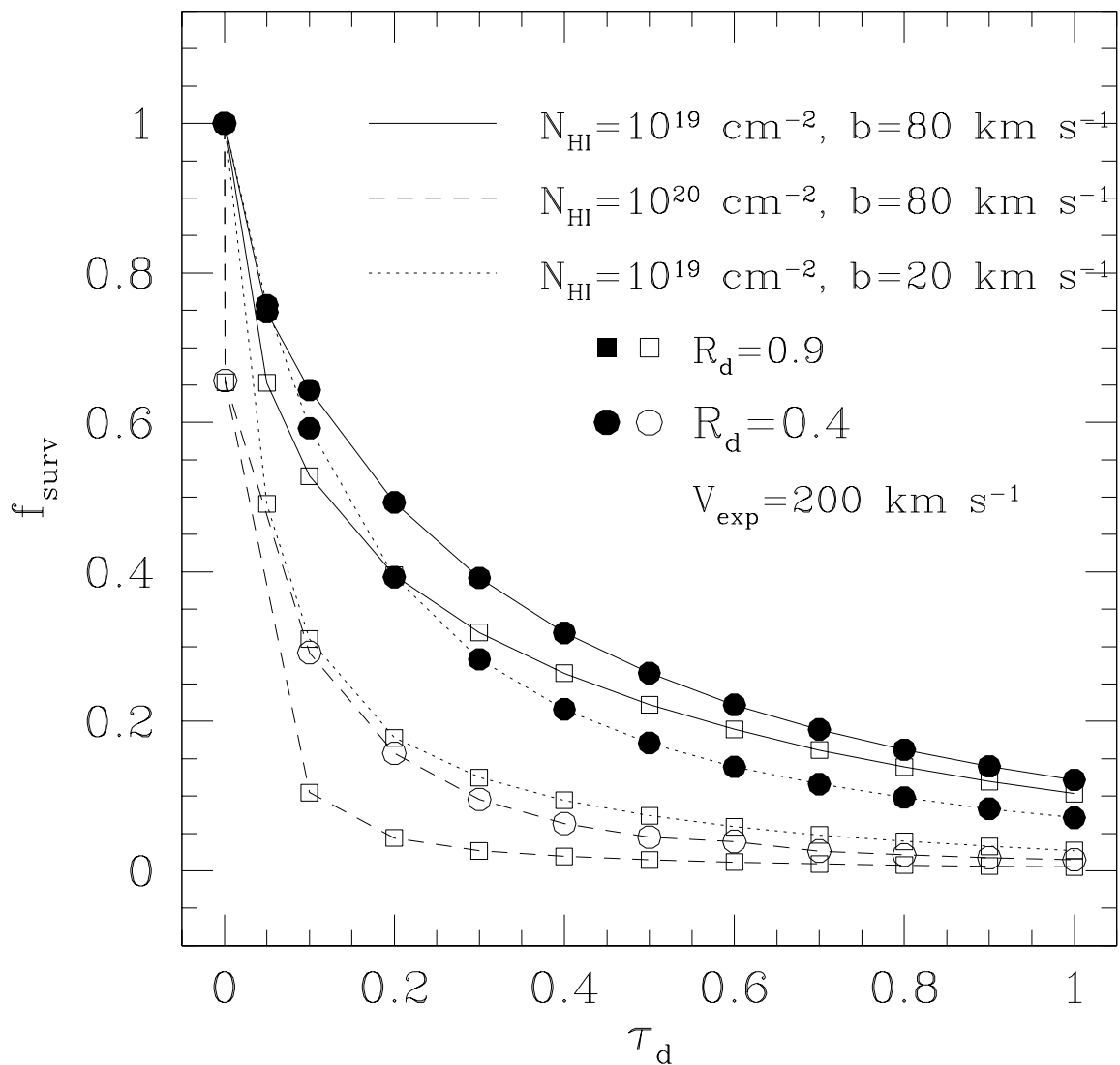}
\caption{Survival fraction of Ly$\alpha$ photons with dust opacities. 
The model parameters of the supershell is given in the figure.
We can see that more Ly$\alpha$ photons are destroyed as $R_d$ increases
to reach $R_{min}$. However, we see in Figures 1 and 2 
that the secondary peak is more rapidly decreased as $R_d$ decreases.}
\label{fig3}
\end{figure}
%%%%%%

\end{document}